УДК 372.8::378.4+004.77

**Маркова О. М.**
ДВНЗ «Криворізький національний університет»

**Моделі використання хмарних технологій у підготовці ІТ-фахівців**

**Анотація.** Стаття присвячена обґрунтуванню доцільності використання хмарних технологій у навчанні математичної інформатики студентів технічних університетів. Мета дослідження, охарактеризованого в статті – аналіз вітчизняного та зарубіжного досвіду використання хмаро орієнтованих засобів ІКТ у підготовці майбутніх фахівців з інформаційних технологій. На основі розглянутого досвіду та проведеного порівняння засобів технологій дистанційного навчання та хмарних технологій визначено переваги використання хмарних технологій для різних категорій учасників навчального процесу та моделі надання хмарних послуг, які доцільно використовувати у процесі навчання нормативних навчальних дисциплін циклів математичної, природничо-наукової й професійної та практичної підготовки майбутніх фахівців з інформаційних технологій.

**Ключові слова:** ІТ-фахівці, хмарні технології, моделі надання хмарних послуг.

**Постановка проблеми.** Однією з необхідних умов фундаменталізації інформатичної освіти у вищих технічних навчальних закладах є переорієнтація базової інформатичної підготовки з опанування швидкозмінними технологіями на стабільні наукові основи інформатики. Провідну роль у розв'язуванні відповідних проблем відіграють комп'ютерне моделювання та обчислювальний експеримент, що одночасно виступають і як методологічна основа інформатики, і як методи навчання інформатичних дисциплін. У роботах М. І. Жалдака [10; 11], Ю. В. Триуса [16] та Т. П. Кобильника [15] показано, що однією із складових фундаменталізації інформатичної підготовки студентів є навчання математичної інформатики, яка, з одного боку, є складовою теоретичної інформатики, де математичні моделі і засоби використовуються для моделювання та дослідження інформаційних процесів у різних сферах діяльності людини, а, з іншого боку базується на використанні інформаційних систем і технологій для розв'язування прикладних задач.

Професійна підготовка студентів технічних університетів, які у майбутній професійній діяльності послугуватимуться моделями та методами математичної інформатики – майбутніх ІТ-фахівців – у ВНЗ України відбувається у галузях знань 12 «Інформаційні технології» (спеціальності 121 «Інженерія програмного забезпечення», 122 «Комп'ютерні науки та інформаційні технології», 123 «Комп'ютерна інженерія», 124 «Системний аналіз», 125 «Кібербезпека») та 15 «Автоматизація та приладобудування» (спеціальність 151 «Автоматизація та комп'ютерно інтегровані технології»). Таке різноманіття призводить до того, що за різними спеціальностями окремі розділи математичної інформатики – фундаментальної основи для всіх інформатичних спеціальностей – відносяться до різних навчальних дисциплін, що читаються на різних курсах і, як правило, не пов'язані одна з одною.

Для подолання цього протиріччя необхідним є визначення цілей та змісту навчання основ математичної інформатики студентів технічних університетів, а також розробка відповідної методики навчання, спрямованої на реалізацію міжпредметних зв'язків та системного підходу в підготовці майбутніх фахівців у галузі ІКТ. Перспективним напрямом навчання таких фахівців є мереже орієнтований підхід, за якого засоби навчання математичної інформатики переносяться у хмарне середовище. За такого підходу в процесі підготовки майбутніх фахівців у галузі ІКТ за спеціальністю 123 «Комп'ютерна інженерія», основним об'єктом діяльності яких є комп'ютерні системи та мережі, хмарні технології стають не лише об'єктом вивчення, а й засобом навчання.

**Аналіз останніх досліджень і публікацій.** Вибір мережних технологій як засобів навчання ґрунтується на дослідженнях Л. В. Брескіної [8], М. Ю. Кадемії [14], Н. В. Морзе [17], Ю. С. Рамського [19], С. В. Шокалюк [30] та інших дослідників. Методично обґрунтоване використання мережних технологій у процесі навчання основ математичної інформатики студентів ІТ-спеціальностей технічних університетів сприятиме організації індивідуальної та колективної навчальної діяльності студентів з використанням хмарних технологій з метою активного, усвідомленого опанування моделями та методами інформаційного моделювання.

Технологічні зміни в Інтернет привели до появи мережних (насамперед соціальних) сервісів Web 2.0, що надає можливості використання у процесі навчання відкритих, безкоштовних і вільних електронних ресурсів, самостійного створення мережних навчальних матеріалів, формування навчальних спільнот тощо [24]. Еволюція та конвергенція Web-технологій привели до появи концепції хмарних обчислень (cloud computing) та відповідних технологій для підтримки навчання та наукових досліджень, насамперед – хмарних Web-СКМ.

Дослідженню стану та перспектив використання хмарних технологій у навчанні присвячені роботи Г. А. Алексаняна [6], В. Ю. Бикова [25], І. С. Войтовича [22], О. О. Жугастрова [12], В. П. Іваннікова [13], О. М. Спіріна [29], Ю. В. Триуса [26] та інших дослідників. Методика використання хмарних технологій у навчанні інформатичних дисциплін сьогодні знаходиться у процесі становлення, тому дослідження можливостей їх використання у процесі навчання основ математичної інформатики студентів технічних університетів є актуальною науково-практичною задачею.

**Формулювання мети написання статті.** Актуальність вище наведених проблем, їх недостатня розробленість в теорії та практиці навчання у вищій технічній школі зумовили вибір мети дослідження – теоретичне обґрунтування та розробка окремих компонентів методичної системи навчання основ математичної інформатики студентів технічних університетів з використанням хмарних технологій, для досягнення якої необхідним є дослідження можливостей застосування засобів хмарних технологій у навчанні інформатичних дисциплін, конкретизоване в меті дослідження – *проаналізувати вітчизняний та зарубіжний досвід використання хмаро орієнтованих засобів ІКТ у підготовці майбутніх фахівців з інформаційних технологій*.

**Подання основного матеріалу дослідження.** К. В. Болгова, розглядаючи використання хмарних обчислень для розробки віртуальних лабораторій, де використовуються методи чисельного моделювання для відтворення досліджуваних процесів і явищ в разі неможливості доступу до реального експериментального устаткування, розглядає наступні моделі автоматизації освітніх процесів:

– IaaS: характеризується віртуалізацією обчислювальної інфраструктури ВНЗ з подальшим її наданням різним підрозділам для виконання власних завдань (у тому числі встановлення спеціалізованих програм для розгортання електронних освітніх ресурсів з доступом через Інтернет);

– PaaS: орієнтована на надання віртуальних ресурсів з уже встановленими обчислювальними пакетами, за допомогою яких забезпечується моделювання та доступ до даних;

– SaaS: традиційна модель надання доступу до програмного забезпечення як до веб-додатку, що забезпечує можливість використання електронних освітніх ресурсів;

– Data as a Service: допоміжна модель, орієнтована на використання хмарних сховищ для колективного доступу до масивів даних, що застосовуються під час роботи з електронними освітніми ресурсами;

– HaaS (Hardware as a Service): специфічна модель для організації віртуальних лабораторій на основі не тільки комп'ютерного моделювання, а й віддаленого доступу до реальних інформаційно-вимірювальних систем або інших технічних засобів [7, с. 6-7].

Дослідник зазначає, що у світлі розвитку мультидисциплінарних напрямів наукових досліджень і відповідних освітніх програм окремої уваги заслуговує модель організації хмарних обчислень Application as a Service, на основі якої забезпечується розробка і використання композитних програм – сукупності взаємопов'язаних хмарних сервісів, орієнтованих на виконання загального завдання. Такі програми створюються за рахунок зв'язування існуючих програмних пакетів у складний програмний комплекс, що виконується на розподілених ресурсах хмарного середовища. Однією з найбільш прогресивних технологій роботи з композитними програмами є концепція, орієнтована на розвиток інтелектуальних технологій підтримки життєвого циклу проблемно-орієнтованих середовищ розподілених обчислень з використанням цієї моделі. Зокрема, на її основі забезпечується інтелектуальна підтримка робочих процесів користувача на основі експертних знань, відчужуваних безпосередньо від розробників предметно-орієнтованих сервісів, навколо яких формується віртуальне професійне співтовариство.

Характеризуючи таке предметно-орієнтоване хмарне середовище, М. Є. Федосін вказує на три типи технологій забезпечення дослідницької діяльності: 1) грід, що надають уніфікований доступ до обчислювального обладнання як до єдиної платформи; 2) хмарні обчислення; 3) веб-лабораторії на основі технологій Web 2.0, спрямовані на організацію предметно-орієнтованих наукових співтовариств з наданням користувачам розвинених засобів комунікації та взаємодії [27, с. 8].

Основна відмінність між грід та хмарними обчисленнями, на думку Є. М. Найнга, полягає у їх використанні: грід переважно використовуються для розв'язування задач за обмежений проміжок часу, а хмарні обчислення орієнтовані на надання довготривалої послуги [9, с. 7]. Дослідник робить висновок про те, що грід та хмарні обчислення доповнюють одне одного.

Г. А. Алексанян вказує, що застосування хмарних технологій дозволяє більш ефективно організовувати самостійну діяльність за рахунок мобільності, доступності й зручності використання на будь-якому пристрої з доступом до Інтернет [6, с. 5]. Впровадження хмарних технологій у навчальний процес вищої та середньої школи, на думку автора, забезпечує [6, с. 45-46]:

– ефективне використання навчальних площ (відпадає необхідність надання окремих та спеціально обладнаних приміщень під традиційні комп'ютерні аудиторії);

– можливість швидкого створення, адаптування і тиражування електронних освітніх ресурсів;

– кардинальне скорочення витрат, необхідних на створення і підтримку комп'ютерних аудиторій;

– можливість для студентів здійснювати зворотний зв'язок із викладачем шляхом оцінювання та коментування пропонованих їм освітніх сервісів;

– мобільність студентів через можливість навчатися в будь-який час і в будь-якому місці, де є доступ до Інтернет;

– гарантію ліцензійності програмного забезпечення, використовуваного у процесі навчання, та скорочення витрат шляхом створення функціонально еквівалентних освітніх сервісів на базі програмного забезпечення з відкритим кодом;

– мінімізацію кількості необхідних ліцензій за рахунок їх централізованого використання;

– можливість централізованого адміністрування програмних та інформаційних ресурсів, що використовуються у навчальному процесі.

Е. А. Альдахіль основою застосування хмарних технологій у навчанні вважає дистанційне навчання [1, с. 21], під яким у вітчизняній нормативній базі «розуміється індивідуалізований процес набуття людиною знань, умінь, навичок і способів пізнавальної діяльності, який відбувається в основному за опосередкованої взаємодії віддалених один від одного учасників навчального процесу у спеціалізованому середовищі, яке функціонує на базі сучасних психолого-педагогічних та інформаційно-комунікаційних технологій» [18]. Основною метою дистанційного навчання є надання освітніх послуг шляхом застосування у навчанні сучасних ІКТ за певними освітніми або освітньо-кваліфікаційними рівнями відповідно до державних стандартів освіти.

У роботі Т. М. Шалкіної [28] розглянуто інноваційне інформаційно-предметне середовище підготовки майбутніх інженерів-програмістів як сукупність педагогічних, інформаційно-комунікативних, матеріально-технічних компонентів, необхідних для організації навчальної діяльності студентів з формування професійних знань і умінь у відповідній предметній галузі в процесі розв'язування професійно-орієнтованих задач. Дослідник наголошує на тому, що інформаційно-предметне середовище стає фактором підготовки майбутніх інженерів-програмістів за реалізації наступних умов:

– наявність автоматизованого навчального середовища і сукупності навчально-методичних матеріалів для організації навчальної діяльності студентів;

– включення студентів у професійну діяльність шляхом запровадження в навчальний процес професійно-орієнтованих задач;

– орієнтація студента на самостійну, пошукову, науково-дослідницьку діяльність з використанням ресурсів мережі Інтернет, різних електронних і друкованих носіїв для пошуку матеріалів стосовно відповідної предметної галузі, аналізу отриманих відомостей, обґрунтування вибору засобів розв'язування задачі.

Остання умова розвинена А. О. Ричковою [20] з позицій особистісно-діяльнісного підходу у напрямі формування професійної самостійності майбутніх інженерів-програмістів на основі дистанційних технологій навчання. Зокрема, дослідником визначені дидактичні, психолого-педагогічні і організаційно-комунікативні можливості використання дистанційних технологій навчання та розроблена відповідна методика.

У хмаро орієнтованому навчальному середовищі дистанційне навчання може бути реалізоване у віртуальному класі (середовищі), в якому навчальна взаємодія студентів та викладача опосередковується хмарними ІКТ. Е. А. Альдахіль зазначає, що із розвитком Web 2.0 все більшої важливості у віртуальному навчальному середовищі набуває соціальна взаємодія [1, с. 22] та вказує на два основні типи віртуальних класів, що розрізняються за режимом взаємодії суб'єктів навчання – асинхронні та синхронні. Проте у більшості випадків доцільними є не їх розмежування, а їх комбінування: так, А. М. Стрюк вказує, що, якщо розглядати форми організації навчання як синхронні (спільна спеціально організована навчальна діяльність у визначений час у визначеному місці – наприклад, у традиційному аудиторному навчанні) та асинхронні (індивідуальна навчальна діяльність, що має бути виконана за певний час – наприклад, у традиційному заочному навчанні), то за всіх форм організації навчального процесу, в тому числі і традиційних комбінування форм організації навчання стає необхідним. Найвищий ступінь асинхронності (та, відповідно, найбільша частина самостійної роботи) має місце за дистанційного навчання [23, с. 20].

Аналізуючи вимоги до системи підтримки комбінованого навчання системного програмування бакалаврів програмної інженерії, А. М. Стрюк у якості важливої основи організації спільної роботи викладача та студентів вказує на віртуальні класи та хмарні технології [23, с. 100]. Навчальне середовище комбінованого навчання Е. А. Альдахіль відносить до хмаро орієнтованих, пропонуючи для побудови такого середовища застосовувати три моделі надання хмарних послуг: SaaS, IaaS, PaaS [1, с. 31].

З. С. Сейдаметова, розглядаючи методичну систему рівневої підготовки інженерів-програмістів [21], відзначає необхідність фундаменталізації навчання дисциплін блоку професійно-орієнтованої та практичної підготовки з одночасним «професійним тюнінгом». Значне місце у цьому, на думку автора, має посідати «люб'язне» професійне програмно-методичне оточення, що є складовою хмаро орієнтованого середовища вищого навчального закладу.

У табл. 1 наведено порівняння засобів технологій дистанційного навчання та засобів хмарних технологій.

*Таблиця 1*
**Порівняння засобів технологій дистанційного навчання та хмарних технологій**

| Параметр | Засоби технологій дистанційного навчання | Засоби хмарних технологій |
|---|---|---|
| Безпека даних | Висока | Низька або середня |

| Параметр | Засоби технологій дистанційного навчання | Засоби хмарних технологій |
|---|---|---|
| Безпека програмного забезпечення | Залежить від навчального закладу | Залежить від постачальника хмарних послуг |
| Вартість апаратного та програмного забезпечення | Середня або висока | Низька |
| Відновлення після збоїв | Ручне | Автоматичне |
| Доступність системи за збою одного з серверів | Відсутня | Наявна |
| Доступність системи незалежно від часових або просторових обмежень | Так | Так |
| Збереження даних та виконання обчислень на комп'ютері користувача | Обов'язково | Не обов'язково |
| Масштабованість навчальних курсів | Обмежена | Необмежена |
| Мобільність програмного забезпечення | Так | Так |
| Можливість інтеграції різних платформ | Низька | Висока |
| Потреба в обслуговуванні системи в разі зміни апаратного чи програмного забезпечення | Наявна | Відсутня |
| Раціональне використання ресурсів | Не завжди | Завжди |
| Управління змістом навчання | Вимагає певних зусиль | Вимагає незначних зусиль |
| Фізичне розташування навчальних матеріалів | Відоме явно | Приховане |

Використання хмарних технологій у підготовці ІТ-фахівців надає ряд переваг:

– *для студентів*: повсюдна доступність необхідних електронних освітніх ресурсів; мобільність програм та даних; відсутність суттєвих програмно-апаратних обмежень на використовувані ресурси; опанування хмарних технологій як провідних для ІТ-галузі; відсутність необхідності адміністрування програмного забезпечення для досягнення найвищої продуктивності під час використання систем програмування та ін.; можливість проведення неруйнівних експериментів у віртуалізованому програмно-апаратному середовищі;

– *для викладачів та співробітників*: можливість використання еластичних хмаро орієнтованих ресурсів (зокрема, з метою розробки завдань різного рівня складності та ресурсоспоживання); можливість уніфікації програмного забезпечення у Web-орієнтованих операційних системах; зниження витрат на адміністрування та утримання ІТ-інфраструктури; широкі можливості моделювання комп'ютерних систем та мереж; можливість простого зберігання та багаторазового використання віртуальних лабораторій;

– *для адміністраторів комп'ютерних систем та мереж*: зниження витрат на ліцензування програмного забезпечення; зняття обмежень на використовуване апаратне та програмне забезпечення завдяки технології віртуалізації; можливість обслуговування потенційно необмеженої кількості студентів; спрощення та уніфікація технічного обслуговування у хмарі.

Таким чином, для студентів хмарні технології виступають насамперед засобом підвищення їхніх професійних компетентностей, для викладачів та співробітників – засобом підвищення ефективності навчання без додаткового навантаження на них, для адміністраторів комп'ютерних систем та мереж – засобом уніфікації та спрощення задач адміністрування. Для ВНЗ у цілому використання хмарних технологій приводить до вагомого зниження витрат на обслуговування, оновлення програмного забезпечення та ліцензування, апаратне конфігурування, забезпечення потужності і зменшення площі інфраструктури, що надає можливість зменшити видатки, не впливаючи на якість надання освітніх послуг.

Переваги використання хмарних технологій для різних категорій учасників навчального процесу узагальнено на рис. 1.

Водночас використання хмарних технологій накладає ряд обмежень.

У першу чергу, це *забезпечення конфіденційності та безпеки*: постачальники хмарних послуг надають їх через центри опрацювання даних, розташовані в різних країнах, у яких може бути законодавчо обумовлена можливість доступу до приватних даних користувачів.

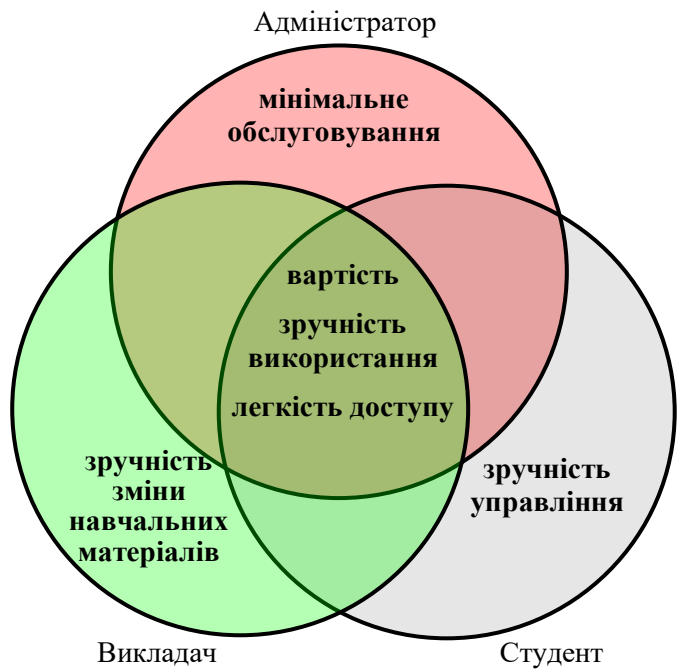

*Рис. 1. Переваги використання хмарних технологій для різних категорій учасників навчального процесу*

Крім того, в разі зберігання даних у публічній хмарі користувач не має фізичного доступу до серверів із своїми даними, і до них потенційно може бути наданий несанкціонований доступ. Для підвищення безпеки постачальники хмарних послуг використовують різні методи. Наприклад, в системі Google визначаються такі напрями забезпечення конфіденційності та безпеки даних: корпоративна культура (перевірка анкетних даних співробітників, навчання всіх співробітників заходів безпеки, інформаційні заходи про безпеку та конфіденційність, окремі підрозділи для забезпечення конфіденційності та безпеки, внутрішній аудит дотримання безпеки, співпраця із фахівцями у галузі безпеки); експлуатаційна безпека (управління вразливостями, запобігання виконання шкідливого програмного забезпечення, моніторинг, розслідування інцидентів); технологічна безпека (новітні центри опрацювання даних, спеціальне серверне апаратне та програмне забезпечення, моніторинг обладнання, захищена глобальна мережа, захист даних під час передавання, незначна затримка та постійна доступність обладнання і послуг); незалежна сторонна сертифікація; домовленість про використання даних (даними володіють клієнти хмарної платформи, дані не скануються та не передаються третім особам, дані фізично вилучаються постачальником протягом 180 днів після вилучення клієнтом); обмеження доступу до даних (логічна ізоляція даних користувачів хмарної платформи, аудит авторизації та доступу, розподіл прав доступу користувача, доступ правоохоронних органів до даних клієнта лише за запитом, контроль дотримання безпеки та конфіденційності з боку сторонніх постачальників хмарних послуг) [4].

Друге обмеження – це *швидкість передавання даних через мережу*: оскільки дані передаються через Інтернет-з'єднання, важливо обрати відповідного постачальника послуг Інтернет та обладнання, щоб уникнути затримки передавання даних. До основних шляхів зменшення мережних затримок відносяться: зменшення затримок на кожному вузлі мережі; мінімізація кількості проміжних мережевих вузлів; усунення перевантаження мережі; зменшення затримок, пов'язаних із транспортним протоколом [2, с. 2].

Третім обмеженням є *сумісність*: кожен постачальник хмарних послуг має власну унікальну архітектуру або реалізацію. Наприклад, в Google використовується для зберігання даних BigTable – розподілена база даних на основі Google File System, що не поширюється та не використовується за межами Google, в той час як в Facebook використовується Apache Cassandra – відкрита розподілена система управління базами даних, що використовується такими компаніями, як Cisco, IBM, Cloudkick, Reddit, Digg, Rackspace і Twitter.

Четверте обмеження – *прив'язування до постачальника хмарних послуг* – виникає за умови використання ним несумісних фірмових форматів даних, обміну повідомленнями тощо. Для запобігання такого прив'язування Дж. Маккендрик рекомендує використовувати хмарні послуги на основі промислових стандартів, а не на основі фірмових технологій постачальника послуг [5]:

1) міжплатформенні інтерфейси на основі WS-I (насамперед HTTP та JSON, включаючи засоби аутентифікації і управління доступом, такі як OAuth), через які забезпечується основна частина реальної мобільності хмарних послуг;

2) програмно-платформенні інтерфейси повинні бути незалежними від постачальника хмарних послуг, специфікації стандарту хмарної платформи для виконання програм повинні бути незалежним від мови, а самі інтерфейси мають бути розроблені для найбільш широко використовуваних мов програмування;

3) специфікація послуг має бути орієнтована на сприйняття людиною, охоплювати всі функціональні і нефункціональні характеристики, підтримувати WS-I разом із WSDL;

4) інтерфейси управління послугами мають бути стандартизовані для масового поширення хмарних послуг як доступного комерційного продукту, тому їх необхідно модифікувати відповідно до нових стандартів, що розробляються DMTF, OASIS, OGF, SNIA та OpenStack;

5) моделі даних мають бути чітко визначені з використанням стандартів форматування тексту (таких, як XML, але не лише за його допомогою) та опису схем;

6) слабкий зв'язок між компонентами є більш доцільним, ніж сильний: сильно пов'язані компоненти важко інтегрувати, особливо у процесі розвитку хмарної системи;

7) орієнтування на обслуговування за відповідною стандартною моделлю надання послуг (IaaS, PaaS, SaaS тощо) надає можливість автоматично враховувати рекомендації щодо слабкого зв'язку між компонентами, специфікації послуг та інтерфейсів;

8) розповсюдження програмного забезпечення через відповідні Інтернет-магазини має бути стандартизоване та налаштовуване відповідно до вимог законодавства;

9) REST-орієнтований доступ до інформаційних ресурсів, зокрема через виклики за протоколом HTTP (GET, POST, PUT, DELETE тощо) має використовуватись замість віддаленого виклику процедур (за RPC-подібними протоколами);

10) формати образів віртуальних машин повинні задовольняти вимоги відкритого формату віртуалізації (OVF DMTF) – це надає можливість переносити образи віртуальних машин між різними хмарними інфраструктурами як одного постачальника хмарних послуг, так і різних.

П'яте обмеження – *ясність угод про рівень обслуговування*, у тому числі щодо таких питань, як гарантована безвідмовна робота і зберігання даних, якщо користувач вирішить змінити постачальника хмарних послуг. Європейською Комісією у межах ініціативи «Європа-2020» розроблені рекомендації щодо стандартизації таких угод, що включають відповідні принципи: нейтральність до технології та бізнес-моделі, повсюдна застосовність, однозначні визначення, порівнювані цілі рівня обслуговування, забезпечення відповідності, наявність стандартів та настанов для різних типів клієнтів, визначення суттєвих характеристик хмарних послуг, доказовість ефективності послуг та ін. [3, с. 5-9].

Шосте обмеження – *масштабованість зберігання* – може бути розв'язане за допомогою концепції еластичності (надання додаткових ресурсів на вимогу), проте на сьогодні в автоматичному режимі це поки що неможливо.

Проведений аналіз освітньо-професійних програм підготовки фахівців з інформаційних технологій в Україні надав можливість визначити моделі надання хмарних послуг, які доцільно використовувати у процесі навчання нормативних навчальних дисциплін (табл. 2). У табл. 2 вказано найменший рівень моделі розгортання, який може бути використаний в процесі навчання відповідної дисципліни.

*Таблиця 2*

**Моделі надання хмарних послуг, які доцільно використовувати у процесі навчання нормативних навчальних дисциплін циклів математичної, природничо-наукової й професійної та практичної підготовки майбутніх фахівців з інформаційних технологій**

| Модель | Навчальні дисципліни |
|---|---|
| SaaS | Вища математика |
| | Теорія ймовірності та математична статистика |
| | Алгоритми та методи обчислень |
| | Дискретна математика |
| | Екологія |
| | Комп'ютерна логіка |
| | Організація баз даних |
| PaaS | Фізика |
| | Теорія електричних та магнітних кіл |
| | Комп'ютерна електроніка |
| | Програмування |
| | Комп'ютерна схемотехніка |
| | Паралельні та розподілені обчислення |
| | Інженерія програмного забезпечення |

| Модель | Навчальні дисципліни |
|---|---|
| IaaS | Архітектура комп'ютерів |
| | Системне програмування |
| | Системне програмне забезпечення |
| | Технології проектування комп'ютерних систем |
| | Комп'ютерні системи |
| | Комп'ютерні мережі |
| | Захист даних у комп'ютерних системах |

**Висновки і перспективи подальших досліджень.**

1. Хмарні технології (хмарні ІКТ) як різновид ІКТ можна визначити як сукупність методів, засобів і прийомів, використовуваних для збирання, систематизації, зберігання та опрацювання на віддалених серверах, передавання через мережу та подання через клієнтську програму всеможливих повідомлень і даних. Витоки хмарних технологій навчання містяться у застосуванні концепції комп'ютерних послуг до навчального процесу, зокрема, надання місця для зберігання електронних освітніх ресурсів та мобільного доступу до них.

2. Як показують результати вивчення досвіду використання хмарних технологій у підготовці ІТ-фахівців, доцільним є застосування у навчанні інформатичних дисциплін таких моделей надання хмарних послуг: «програмне забезпечення як послуга», «платформа як послуга» та «інфраструктура як послуга» на основі інформатичної технології віртуальних машин та педагогічної технології дистанційного навчання. Однією з явних переваг використання хмарних технологій у підготовці майбутніх ІТ-фахівців в технічних університетах є можливість використання сучасних засобів паралельного програмування як основи хмарних технологій.

**Список використаних джерел**

1. Aldakheel E. A. A Cloud Computing framework for computer science education : A Thesis Submitted to the Graduate College of Bowling Green State University in partial fulfillment of the requirements for the degree of Master of Science / Eman A. Aldakheel ; the Graduate College of Bowling Green State University. – [Bowling Green] : December, 2011. – XI, 130 p.

2. Architecting Low Latency Cloud Networks : Arista Whitepaper [Electronic resource] / Arista. – [2009-05-08]. – 5 p. – Access mode : https://www.arista.com/assets/data/pdf/CloudNetworkLatency.pdf

3. Cloud Service Level Agreement Standardisation Guidelines [Electronic resource]. – Brussels. – 24/06/2014. – 41 p. – Access mode : http://ec.europa.eu/information_society/newsroom/cf/dae/document.cfm?action=display&doc_id=6138

4. Google Security Whitepaper [Electronic resource] // Google Cloud Platform. – May 26, 2015. – Access mode : https://cloud.google.com/security/whitepaper

5. McKendrick J. 10 steps to avoid cloud vendor lock-in [Electronic resource] / Joe McKendrick // ZDNet. – July 12, 2013. – Access mode : http://www.zdnet.com/article/10-steps-to-avoid-cloud-vendor-lock-in/

6. Алексанян Г. А. Формирование самостоятельной деятельности студентов СПО в обучении математике с использованием облачных технологий : дисс. ... канд. пед. наук : 13.00.02 – теория и методика обучения и воспитания (математика) / Алексанян Георгий Ашотович ; Федерал. гос. бюдж. образов. учрежд. ВПО «Армавирская государственная педагогическая академия». – Армавир, 2014. – 150 с.

7. Болгова Е. В. Автоматизация процесса разработки виртуальных лабораторий на основе облачных вычислений : автореф. дисс. ... канд. техн. наук : 05.13.06 – автоматизация и управление технологическими процессами и производствами (образование) / Болгова Екатерина Владимировна ; [Санкт-Петербургский нац. исслед. ун-т инф. технол., механики]. – С-Пб., 2012. – 18 с.

8. Брескіна Л. В. Професійна підготовка майбутніх вчителів інформатики на основі сучасних мережевих інформаційних технологій : автореф. дис... канд. пед. наук : 13.00.02 – теорія та методика навчання інформатики / Брескіна Лада Валентинівна ; Нац. пед. ун-т ім. М. П. Драгоманова. – К., 2003. – 17 с.

9. Е Мьинт Найнг. Разработка системы запуска ресурсоемких приложений в облачной гетерогенной среде : автореф. дисс. ... канд. техн. наук : 05.13.15 – вычислительные машины, комплексы и компьютерные сети / Е Мьинт Найнг ; [Санкт-Петербургский гос. электротехн. ун-т «ЛЭТИ» им. В. И. Ульянова (Ленина)]. – С-Пб., 2013. – 19 с.

10. Жалдак М. И. О некоторых методических аспектах обучения информатике в школе и педагогическом университете / М. И. Жалдак // Методология и технология образования в XXI веке: математика, информатика, физика : материалы международной научно-практической конференции 17-18 ноября 2005 г. / Министерство образования республики Беларусь ; Учреждение образования

«Белорусский государственный педагогический университет имени Максима Танка». – Минск, 2006. – С. 260-268.

11. Жалдак М. І. Основи теорії і методів оптимізації : [навч. посіб. для студ. мат. спец. вищ. навч. закл.] / М. І. Жалдак, Ю. В. Триус. – Черкаси : Брама-Україна, 2005. – 607 с.

12. Жугастров О. О. Хмарні обчислення: сутність, недоліки, переваги / О. О. Жугастров // Комп'ютер у школі та сім'ї. – 2011. – № 2. – С. 54-56.

13. Иванников В. П. Облачные вычисления в образовании, науке и госсекторе [Электронный ресурс] / Иванников Виктор Петрович ; Институт системного программирования Российской Академии Наук. – [М.], [2010]. – 23 с. – Режим доступа : http://grid2010.jinr.ru/files/pdf/cloud.pdf

14. Кадемія М. Ю. Формування професійних знань учнів профтехучилищ засобами мережних комунікацій: дис. ... канд. пед. наук : 13.00.04 – теорія та методика професійної освіти / Кадемія Майя Юхимівна ; Інститут педагогіки і психології професійної освіти АПН України. – К., 2004. – 260 с.

15. Кобильник Т. П. Методична система навчання математичної інформатики у педагогічному університеті : автореф. дис. ... канд. пед. наук : 13.00.02 – теорія та методика навчання (інформатика) / Кобильник Тарас Петрович ; Нац. пед. ун-т ім. М. П. Драгоманова. – К., 2009. – 19 с.

16. Копаєв О. В. Фундаментальний аспект базового курсу інформатики / Копаєв О. В., Триус Ю. В. // Сучасний стан і перспективи шкільних курсів математики та інформатики у зв'язку з реформуванням у галузі освіти (Дрогобич, 14–16 листопада 2000 р.) : [всеукраїнська науково-практична конференція] : тези доповідей. – Дрогобич : ДДПУ, 2000. – С. 138-140.

17. Морзе Н. В. Система методичної підготовки майбутніх вчителів інформатики в педагогічних університетах : дис. ... д-ра пед. наук : 13.00.02 – теорія і методика навчання інформатики / Морзе Наталія Вікторівна ; Нац. пед. ун-т ім. М. П. Драгоманова. – К., 2003. – 605 с.

18. Про затвердження Положення про дистанційне навчання [Електронний ресурс] / МОН України : Наказ № 466, Положення. – 25.04.2013. – Режим доступу : http://zakon4.rada.gov.ua/laws/show/z0703-13

19. Рамський Ю. С. Вивчення інформаційно-пошукових систем мережі Інтернет : навч.-метод. посіб. / Ю. С. Рамський, О. В. Рєзіна ; Нац. пед. ун-т ім. М. П. Драгоманова. – К. : [б. в.], 2004. – 60 с.

20. Рычкова А. А. Дистанционные образовательные технологии как средство формирования профессиональной самостоятельности будущих инженеров-программистов : диссертация ... кандидата педагогических наук : 13.00.08 – теория и методика профессионального образования / Рычкова Анастасия Александровна ; [Оренбург. гос. ун-т]. – Оренбург, 2010. – 235 с.

21. Сейдаметова З. С. Методична система рівневої підготовки майбутніх інженерів-програмістів за спеціальністю «Інформатика» : автореф. дис... д-ра пед. наук : 13.00.02 – теорія та методика навчання (інформатика) / Сейдаметова Зарема Сейдаліївна ; Нац. пед. ун-т ім. М. П. Драгоманова. – К., 2007. – 40 с.

22. Сергієнко В. П. Створення навчальних ресурсів у середовищі Moodle на основі технології „Cloud computing" [Електронний ресурс] / Сергієнко Володимир Петрович, Войтович Ігор Станіславович // Інформаційні технології і засоби навчання. – 2011. – Том 24, №4. – 8 с. – Режим доступу : http://journal.iitta.gov.ua/index.php/itlt/article/download/518/434

23. Стрюк А. М. Система «Агапа» як засіб навчання системного програмування бакалаврів програмної інженерії : дис. ... канд. пед. наук : 13.00.10 – інформаційно-комунікаційні технології в освіті / Стрюк Андрій Миколайович ; Національна академія педагогічних наук України, Інститут інформаційних технологій і засобів навчання. – К., 2012. – 312 с.

24. Сучасні технології дистанційного навчання / [Кухаренко Володимир Миколайович] ; Дистанційні курси НТУ «ХПІ». – Харків. – Режим доступу : http://dl.kharkiv.edu/course/view.php?id=7

25. Технології хмарних обчислень – провідні інформаційні технології подальшого розвитку інформатизації системи освіти в Україні (Відповіді доктора технічних наук, професора, академіка НАПН України, лауреата Державної премії, заслуженого діяча науки і техніки України, директора Інституту інформаційних технологій і засобів навчання НАПН України Валерія Юхимовича Бикова на запитання головного редактора науково-методичного журналу «Комп'ютер у школі та сім'ї» В.Д. Руденка) // Комп'ютер у школі та сім'ї. – 2011. – № 6(94). – С. 3-11.

26. Триус Ю. В. Інноваційні технології навчання у вищій школі [Електронний ресурс] / Триус Ю. В. ; Черкас. держ. технол. ун-т // X Міжвуз. школа-семінар «Сучасні педагогічні технології в освіті». – Харків, 31.01–02.02.2012. – 52 с. – Режим доступу : http://www.slideshare.net/kvntkf/tryus-innovacai-iktvnz

27. Федосин М. С. Виртуализация многокомпонентной системной архитектуры предметно-ориентированной облачной вычислительной среды : автореф. дисс. ... канд. техн. наук : 05.13.15 – вычислительные машины, комплексы и компьютерные сети / Федосин Михаил Евгеньевич ;

28. Шалкина Т. Н. Информационно-предметная среда как фактор подготовки будущих инженеров-программистов : диссертация ... кандидата педагогических наук : 13.00.08 – теория и методика профессионального образования / Шалкина Татьяна Николаевна ; Оренбургский гос. ун-т. – Оренбург, 2003. – 190 с.

29. Шишкіна М. П. Проблеми інформатизації освіти в Україні в контексті розвитку досліджень оцінювання якості засобів ІКТ [Електронний ресурс] / Шишкіна Марія Павлівна, Спірін Олег Михайлович, Запорожченко Юлія Григорівна // Інформаційні технології і засоби навчання. – 2012. – Том 27, №1. – 17 с. – Режим доступу : http://journal.iitta.gov.ua/index.php/itlt/article/download/632/483

30. Шокалюк С. В. Методичні засади комп'ютеризації самостійної роботи старшокласників у процесі вивчення програмного забезпечення математичного призначення : автореф. дис. ... канд. пед. наук : 13.00.02 – теорія і методика навчання (інформатика) / Шокалюк Світлана Вікторівна ; Нац. пед. ун-т ім. М. П. Драгоманова. – К., 2010. – 21 с.

**Модели использования облачных технологий в подготовке ИТ-специалистов**
*Маркова О. Н.*

**Аннотация**. Статья посвящена обоснованию целесообразности использования облачных технологий в обучении математической информатике студентов технических университетов. Цель исследования – анализ отечественного и зарубежного опыта использования облачно ориентированных средств ИКТ в подготовке будущих специалистов в области информационных технологий. На основе рассмотренного опыта и проведенного сравнения средств технологий дистанционного обучения и облачных технологий определены преимущества использования облачных технологий для различных категорий участников учебного процесса и модели предоставления облачных услуг, которые целесообразно использовать в процессе обучения нормативным учебным дисциплинам циклов математической, естественно-научной и профессиональной и практической подготовки будущих ИТ-специалистов.

**Ключевые слова:** ИТ-специалисты, облачные технологии, модели предоставления облачных услуг.

**Models of using cloud technologies at the IT professionals training**
*Markova O. N.*

**Resume**. The article is devoted to the rationale of the use of cloud technologies in teaching mathematical informatics students of technical universities. Purpose of the article – the analysis of domestic and foreign experience in the use of cloud-oriented ICT in the training of future professionals in the field of information technology. Based on a review of experiences and comparisons tools of distance learning technologies and cloud technologies identified the advantages of using cloud technologies for different categories of the learning process participants and models of cloud services, which should be used in training the regulatory academic disciplines cycles of mathematical, scientific and vocational & practical training of the future IT professionals.

**Keywords:** IT professionals, cloud technologies, cloud services delivery models.

94